\documentclass{PoS}

\usepackage[fleqn]{amsmath} 						

\usepackage[all,2cell]{xy} \UseAllTwocells 		

\usepackage[mathscr]{eucal}   						
\newcommand{\As}{{\mathscr{A}}}\newcommand{\Bs}{{\mathscr{B}}}\newcommand{\Cs}{{\mathscr{C}}}

\newcommand{\Es}{{\mathscr{E}}}
\newcommand{\Hs}{{\mathscr{H}}}
 
\newcommand{\Ms}{{\mathscr{M}}}\newcommand{\Ns}{{\mathscr{N}}} \newcommand{\Os}{{\mathscr{O}}}

\newcommand{\Ts}{\mathscr{T}} 			

\newcommand{\Xs}{{\mathscr{X}}}

\newcommand{\CC}{{\mathbb{C}}}
\newcommand{\MM}{{\mathbb{M}}}

\DeclareMathOperator{\Aut}{Aut} 
\DeclareMathOperator{\Sp}{Sp} 

\title{Categorical Operator Algebraic Foundations of Relational Quantum Theory}

\ShortTitle{Categorical Operator Algebraic Foundations of Relational Quantum Theory}

\author{\speaker{Paolo Bertozzini}\thanks{
We acknowledge the partial support from the Department of Mathematics and Statistics in Thammasat University and the Thammasat University Research Grant n. 2/15/2556: ``Categorical Non-commutative Geometry''. 
Thanks also to Starbucks Coffee shops in Emporium Suites Tower, UBC II Tower, Jasmine City in Sukhumvit, where most of the plesant time dedicated to this research project was spent. Thanks to C.Rovelli for providing a copy of the manuscript~\cite{R-half-way}. 
}\\
Department of Mathematics and Statistics, Faculty of Science and Technology, Thammasat University, Khlongneung, Khlongluang, 12121 Pathumthani, Thailand \\
E-mail: \email{paolo.th@gmail.com}}

\abstract{We provide an algebraic formulation of C.Rovelli's relational quantum theory that is based on suitable notions of ``non-commutative'' higher operator categories, originally developed in the study of categorical non-commutative geometry. 
As a way to implement C.Rovelli's original intuition on the relational origin of space-time, in the context of our proposed algebraic approach to quantum gravity via Tomita-Takesaki modular theory, we tentatively suggest to use this categorical formalism in order to spectrally reconstruct non-commutative relational space-time geometries from categories of correlation bimodules between operator algebras of observables.

Parts of this work are joint collaborations with: Dr.Roberto Conti (Sapienza Universit\`a di Roma), Assoc.Prof.Wicharn Lewkeeratiyutkul (Chulalongkorn University, Bangkok), Dr.Rachel Dawe Martins (Istituto Superior T\'ecnico, Lisboa), Dr.Matti Raasakka (Paris 13 University), Dr.Noppakhun Suthichitranont. 
}

\FullConference{Frontiers of Fundamental Physics 14 - FFP14,\\
                15-18 July 2014\\
                Aix Marseille University (AMU) Saint-Charles Campus, Marseille, France}
		
\begin{document}

\section{Relational Quantum Theory - Background} 

Relationalism in physics has a long tradition going back at least to the work of G.Leibniz, G.Berkeley, E.Mach, J.Wheeler, among others. 
Relational dynamics is a core feature of Einstein's theory of relativity (special and general): the dynamics is not specified as an explicit functional evolution with respect to a time parameter, but it is given by an implicit relation between the several variables (Rovelli's partial observables).  
Similarly (Einstein's hole argument), localization of events in general relativity is not absolute: coordinates are gauge and points on a Lorentz manifold are not objective elements of the theory (coincidences, events and correlations, that are preserved by local diffeomorphisms, are). 
In this classical context, mathematically speaking, the transition is between functions and relations (more generally 1-quivers).

In 1994, C.Rovelli~\cite{R-rqm} elaborated \emph{relational quantum mechanics} as an attempt to radically solve the interpretational problems of quantum theory. This approach is based on two assumptions: 1) \emph{relativism}: all systems (necessarily quantum) have equivalent status, there is no difference between observers and objects; 2) \emph{completeness}: quantum physics is a complete and self-consistent theory of natural phenomena. 
An analysis of the Schr\"odinger's cat problem entails: 
a) states are \emph{relative} to each observer: different observers can give different (but ``compatible'') accounts of the interactions; 
b) the only physical properties (interactions) are \emph{correlations} between observers;  
c) physics is about \emph{information} exchange between agents: correlations describe the ``relative information'' that observers posses about each other.  

In 1996 C.Rovelli went even further~\cite{R-half-way} with the radical conjecture stating that there is a direct connection between quantum relationalism via correlations of systems and  the general relativistic relational status of space-time localization determined by contiguity of events. 
This strongly suggests that it should be possible to reinterpret the information on space-time localization (contiguity) as correlations (interactions) between quantum systems, opening the way for a reconstruction of space-time ``a-posteriori'' from purely quantum correlations 
(see also R.Haag~\cite{H}). 

It is our purpose to provide some possible mathematical implementation in support of this approach to quantum relativity and, as a general mathematical framework for relational physics, we propose to formalize ``correlations'' (relations between quantum systems) and their ``compatibility'' using a higher C*-categorical environment: 
a) systems and observers are represented by \hbox{C*-algebraic} data;  
b) correlations and interactions are represented by ``suitable bimodules''; 
c) there is a modular hierarchy of systems in mutual correlation, because we must distinguish ``observers'' from ``observers of observers'', 
``observers of observers of observers'' and so on; 
d) the mutual compatibility requested is encoded by the covariance coming from the compositions operations of a higher category; 
e) systems with higher internal correlations are described by hyper-C*-algebras. 

\section{Quantum Relations} 

Following the framework of \emph{algebraic quantum theory} (see for example F.Strocchi~\cite{S}) we accept as temporary assumption that: 
quantum systems can be described as \emph{C*-algebras}, classical systems, as a special case, being described by commutative C*-algebras. Since  Gel'fand-Na\u\i mark duality assures that every Abelian C*-algebra $\As$ is $*$-isomorphic to the algebra $C(\Sp(\As))$ of continuous functions over its spectrum $\Sp(\As)$ that is a locally compact Hausdorff topological space: 
\textit{classical phase space $\simeq$ spectrum of Abelian C*-algebra $\simeq$ locally compact Hausdorff space}.  

Our work on spectral theory of commutative full C*-categories~\cite{BCL-hcat} motivates this ``Spectral Conjecture'': \emph{quantum spaces 
$\simeq$ non-commutative C*-algebras, can be spectrally described via ``families'' of rank-one Fell bundles over involutive categories}, where, in line with ideas from A.Connes and L.Crane, non-commutativity and quantization can be traced back to non-trivial relations (1-arrows) between points i.e.~to the categorical structure of the spectrum. 

Classical relations (1-arrows between points $\simeq$ 2-arrows between 1-loops), as morphisms of classical spaces, are dually described as bimodules (of sections of line-bundles over a 1-quiver). 
\\
Quantum relations (2-arrows between 1-arrows), as morphisms of quantum spaces (points with relations), will be dually described as higher bimodules of sections of line-bundles over higher quivers: \emph{quantum relations $\simeq$ (higher) bimodules}. 

Exemplifications of this ideology are already abundant in the formalism of quantum theory:  
a) 
inclusions of subsystems (homomorphisms) and symmetries (isomorphisms) $\phi:\As\to\Bs$ give adjoint pairs of twisted bimodules 
${}_\phi\Bs,\Bs_\phi$;  
b) 
states $\omega$ on $\As$, via GNS-representation $(\Hs_\omega,\pi_\omega,\xi_\omega)$, give bimodules ${}_\As(\Hs_\omega)_\CC$; 
c) 
conditional expectations $\Phi:\As\to\Bs$ give $\As$-$\Bs$ bimodules via Kasparov GNS-representation theorem. 

To implement the idea of ``Rovelli's relational network'', we turn to (higher) category theory, whose usage in quantum physics was already pioneered by J.Roberts, L.Crane, J.Baez and others. 

Different observers are now mutually related by a family of quantum correlation channels, some of them describing symmetries, others quantum interactions. 
Each observer is still equipped with a family of potential states, but now states of different observers can be compared via the family of binary correlations so far introduced. 
The dynamic of the quantum theory has been totally codified via correlations and the potentially huge Cartesian product of state-spaces of the observers is now collapsed to a much more manageable set of states that are compatible under the given correlations. 

As a first step in the mathematical formalization of C.Rovelli relational quantum mechanics we propose the following statement: 
\textit{a physical system is totally captured by such a ``category'' of bimodules of binary correlations\footnote{In principle we might also try to consider $n$-tuple correlations between observers. 
Multimodules and their \hbox{C*-polycategories} would be necessary to formalize mathematically such notions (work in progress).} 
(C.Rovelli's ``relational network'')}.
Although a physical system is for now formalized as a 1-categorical structure (level-1 correlations between algebras of observables of different agents), the ``vertical categorification catastrophe'' is almost inescapable: 
a) mathematically, the family of physical systems itself is a 2-category (via functors and natural transformations); 
b) the ideological assumption of role interchangeability between systems and observers requires that such higher categories should be physically relevant: the systems must themselves be observers, object of further correlations; 
c) 
correlations between two systems could in principle be reconducted to lower level correlations between their ``internal agents'', but this 
\emph{reductionist approach} is not compatible with the original introduction of observers as ``black building blocks'' whose internal correlation structure is ``not affected'' by the (several alternative) external quantum correlations! 
Given two quantum systems $\As,\Bs$, a pair of observers can give a different description of their interaction correlations: 
$\As \xrightarrow{\Ms}\Bs$, $\As \xrightarrow{\Ns}\Bs$. 
The mutual compatibility between the two correlations is described by a ``higher level'' morphism in a 2-category: 
{\footnotesize 
\xymatrix{\As \rtwocell^\Ms_\Ns{\Phi}& \Bs.} 
} \\ 
The morphism $\Ms\xrightarrow{\Phi}\Ns$ can be seen as a (level-2) correlation bimodule between the C*-algebroids $\Ts(\Ms)$, $\Ts(\Ns)$ generated respectively by $\Ms$ and $\Ns$. And so on \dots 

This opens the way to the scary possibility to have different levels of ``reality'' for quantum properties, since observers and systems are now not only ``extensively'' related, but ``hierarchically'' structured. \textit{Higher categories will be necessary to formalize this situation.} 

One might propose a \emph{hypercovariance principle} to deal with the invariance of the physics along the hierarchical ladder of observers/systems. 
Higher C*-categories and higher Fell bundles have been developed from the beginning with this kind of goals in mind and can potentially deal with such a context of interacting ``structured virtual realities''. 

In order to mathematically implement this program, two extremely important mathematical obstacles must be overcome: 
1) to describe higher level \emph{relational} situations we need to develop a theory of involutions for higher categories;  
2) due to \emph{Eckmann-Hilton collapse}, usual higher category theory cannot accommodate in a non-trivial way non-commutativity (quantum subsystems)!  

\section{Quantum Higher $*$-Categories}

A strict \emph{globular $n$-category} $(\Cs,\circ_0,\dots \circ_{n-1})$ is a family $\Cs$ of $n$-arrows equipped with partially defined binary compositions $\circ_p$, for $p:=0,\dots, n-1$, such that:  
a) 
for all $p=0,\dots, n-1$, $(\Cs,\circ_p)$ is a 1-category, with $\Cs^p$ denoting its partial identities; 
b)
for all $q<p$, $\Cs^q\subset \Cs^p$ i.e.~a $\circ_q$-identity is also a $\circ_p$-identity; 
c) 
for all $p,q=0,\dots n-1$, with $q<p$, $\Cs^p\circ_q\Cs^p\subset \Cs^p$, i.e.~the $\circ_q$-composition of $\circ_p$-identities, whenever exists, is a 
$\circ_p$-identity; 
d)  
the \emph{exchange property} holds for all $q<p$: whenever $(x\circ_p y)\circ_q (w\circ_p z)$ exists also $(x\circ_q w)\circ_p (y\circ_p z)$ exists and they coincide.\footnote{For semplicity we treat here only the case of strict globular higher categories. Strict cubical 
higher categories~\cite{BCM} and weak $n$-categories might be used as well. 
For higher categorical background we refer to T.Leinster~\cite[chapter~1]{L}.}

Introducing involutions on $n$-(C*)-categories has been relatively straightforward~\cite{BCL-ncg,BCL-cncg,BCLS,BCM}: 
we have an \emph{involutive (higher) category} whenever there are some duality maps $*_\alpha:\Cs\to \Cs$, with index 
$\alpha\subset \{0,\dots,n-1\}$, that are: 
a) 
covariant functors for all $\circ_q$ with $q\notin \alpha$; 
b) 
contravariant functors for all $\circ_q$ with $q\in \alpha$;  
c) 
involutive: $(x^{*_\alpha})^{*_\alpha}=x$; 
d) 
Hermitian: $x^{*_\alpha}=x$, for all $\circ_q$-identities, with $q=\min(\alpha)$; 
e) 
commuting: $(x^{*_\alpha})^{*_\beta}=(x^{*_\beta})^{*_\alpha}$. 
The higher category $(\Cs,\circ_0,\dots,\circ_{n-1})$ is \emph{fully involutive} if its family of involutions generates all possible $2^n$ dualities of 
$n$-arrows. 

The problem of compatibility with non-commutative subsystems is much more delicate, since the exchange property (d) now assumed for 
$n$-categories implies the  Eckmann-Hilton collapse: for $q< p<n$ and $n$-arrows with a common $q$-source $q$-target,  $\circ^n_p=\circ^n_q$ and they are both \emph{commutative} operations! This means that it is perfectly possible to have non-commutative C*-algebras as ``subsystems'' of a 
1-C*-category, but that only classical subsystems can be nodes of correlations at depth higher than 2 in $n$-C*-categories for $n\geq 2$. 
To avoid such trouble, we proposed~\cite{BCLS} the following weakened \emph{non-commutative exchange property}:
for all $\circ_p$-identities $\iota\in \Cs^p$, for all $q<p$, the partially defined maps $\iota\circ_q-:(\Cs,\circ_p)\to(\Cs,\circ_p)$ and 
$-\circ_q \iota:(\Cs,\circ_p)\to(\Cs,\circ_p)$ are functorial. 

Following~\cite{BCLS}, a \emph{quantum strict globular $n$-C*-category} $(\Cs,\circ_0,\dots,\circ_{n-1},*_0,\dots,*_{n-1},+,\cdot,\|\cdot\|)$ is a fully involutive strict $n$-category with \emph{non-commutative exchange} such that: 
a) 
if $a,b\in \Cs^{n-1}$, the fiber $\Cs_{ab}:=\{x\in \Cs \ | \  b\circ_{n-1} x, \ \ x\circ_{n-1} a \ \text{both exist}\}$ is Banach with norm 
$\|\cdot\|$;  
b) 
for all $p$, $\circ_p$ is fiberwise bilinear and $*_p$ is conjugate-linear; 
c) 
for all $\circ_p$, $\|x\circ_p y\|\leq\|x\|\cdot\|y\|$, whenever $x\circ_p y$ exists; 
d) 
for all $p$, $\|x^{*_p}\circ_p x\|=\|x\|^2$, for all $x\in \Cs$; 
e) 
for all $p$, $x^{*_p}\circ_p x$ is positive in $(\Es(\Cs_{ee}),\circ_p,*_p,+,\cdot,\| \cdot \|)$, the C*-algebra envelope of $\Cs_{ee}$, where $e$ is the 
$p$-source of $x$.\footnote{A \emph{partially involutive strict $n$-C*-category} will be equipped with only a subfamily of the previous involutions and will satisfy only those properties that can be formalized using the involutions available.}

Recalling from W.Heisenberg how matrix algebras $\MM_{N\times N}(\CC)$ first appeared in physics, as convolutions algebras for the pair groupoid 
$N\times N$ (matrices being sections of a complex rank-one (Fell) bundle over $N\times N$), abundant examples of the previous definition can be obtained taking any finite globular involutive $n$-category $(\Xs,\circ_0,\dots,\circ_{n-1},*_0,\dots,*_{n-1})$ (with usual or non-commutative exchange) in place of $N\times N$ and any associative unital complex C*-algebra $\As$ in place of $\CC$. 
The family of sections $\MM_\Xs(\As)$ of the bundle $\Es:=\Xs\times \As$ is a \emph{hyper-convolution algebra} with $n$ operations 
$(\sigma\circ_p \rho)_z:=\sum_{x\circ_p y=z}\sigma_x\cdot_{{}_\As} \rho_y$ and $n$ involutions 
$(\sigma^{*_p})_z:=(\sigma_{z^{*_p}})^{*_{{}_\As}}$, for $p=0,\dots,n-1$. $\Es\subset \MM_\Xs(\As)$ becomes a quantum strict globular involutive $n$-C*-category inside $\MM_\Xs(\As)$ and we can think of the sections $\sigma\in \MM_\Xs(\As)$ of $\Es$ as ``hypermatrices'' whose entries $\sigma_x\in\As$ are indexed by $n$-arrows in a globular strict finite involutive $n$-category $\Xs$ in place of the pair groupoid $N\times N$.  

In this way we realize that a quantum system with higher internal correlations might actually be described by a \emph{hyper C*-algebra}: 
a complete topo-linear space $\As$, equipped with different pairs of multiplication/involution $(\circ_k,*_k)$, for $k=0,\dots n-1$, inducing on $\As$ a 
C*-algebra structure, via a necessarily unique C*-norm $\| \cdot\|_k$ compatible with the given topology. 
Apart from the hyper-convolution algebras $\MM_\Xs(\As)$ above, examples of finite hyper-C*-algebras are provided by depth-$n$ hypermatrices 
$\MM_{\Xs_1\times\cdots\times\Xs_n}(\CC):=\MM_{\Xs_1}(\cdots\MM_{\Xs_n}(\CC)\cdots)\simeq 
\MM_{\Xs_1}(\CC)\otimes\cdots\otimes\MM_{\Xs_n}(\CC)$, $\Xs_j:=N_j\times N_j$, equipped with the family of $2^n$ multiplications acting at each depth-level either as convolution or as Schur product and the family of $2^n$ involutions acting either trivially or by adjunction.\footnote{Also these depth-$n$ hypermatrices can be seen~\cite{BCLS} as convolution hyper C*-algebras of \emph{cubical} $n$-categories equipped with $2^n$ compositions \dots\ another hint pointing towards the usefulness of ``non-standard higher categories''.} 

\section{Relational Spectral Space-Time} 

The formalization of relational quantum theory via higher C*-categories is only one of the intermediate steps in our ongoing research program on 
\emph{modular algebraic quantum theory}~\cite{BCL-mqg,BCL-ncg} where: 
a)  \emph{quantum theory}, as a fundamental theory of physics, \emph{does not come from a quantization}; 
b)  \emph{geometry}, as a variant of \emph{A.Connes' non-commutative geometry}~\cite{C}, must be \emph{spectrally reconstructed a posteriori} from a basic operational theory of covariant observables and states, using as basic tool  \emph{Tomita-Takesaki modular theory}~\cite{T}; 
c)  \emph{categories of operational data} provide the general framework for the formulation of covariance \dots\ and ultimately for the identification of the geometric degrees of freedom (space-time) hidden in the theory. 

More specifically~\cite{BCL-mqg,BCL-ncg}, every state $\omega$ on a C*-algebra $\Os$ of partial observables induces a net of subalgebras 
$\As\subset \Os$ such that $\omega |_\As$ is a KMS-state.
By \emph{Tomita-Takesaki theory}~\cite{T}, every such KMS-state $\omega$ on a C*-subalgebra $\As$ uniquely determines a modular spectral 
\emph{non-commutative geometry} $(\As_\omega,\Hs_\omega,\xi_\omega,K_\omega,J_\omega)$ where: 
$\Hs_\omega$ is the Hilbert space of the GNS representation $\pi_\omega$ induced by $\omega|_\As$, with cyclic separating unit vector 
$\xi_\omega\in \Hs_\omega$; the operator $K_\omega:=\log \Delta_\omega$ is the generator of the one-parameter unitary group $t\mapsto \Delta_\omega^{it}$ spatially implementing the modular one-parameter group of $*$-automorphisms $\sigma^\omega_t\in\Aut(\As)$; the operator 
$J_\omega$ is the conjugate-linear operator spatially implementing the modular conjugation anti-isomorphism 
$\gamma_\omega:\pi_\omega(\As)\to \pi_\omega(\As)'$; 
and $\As_\omega:=\{a\in \As \ | \ [K_\omega,\pi_\omega(a)]\in\pi_\omega(\As)''\}$, with $\pi(\As)',\pi(\As)''$ the (bi)commutant of $\As$ in $\Bs(\Hs_\omega)$. 

Tomita-Takesaki modular theory is here taking the role of the quantum version of Einstein's equation associating ``geometries'' to ``matter content'' where: ``geometries'' are spectrally described by variants of \emph{modular spectral triples} 
(see~\cite{CRPS} for references); and ``matter content'' is described by the set of quantum correlations between observables specified by the state. Each pair $(\Os,\omega)$ gives a different ``net'' $(\As_\omega,\Hs_\omega,\xi_\omega,K_\omega,J_\omega)_{\As\subset\Os}$ of modular spectral geometries that are: 
\emph{quantum}, since $\As\subset \Os$ are non-commutative; \emph{state-dependent} on $\omega$; and \emph{relative to observers} $\Os$. 

The pair $(\Os,\omega)$ selects C*-categorical data inside the C*-algebra $\Os$: the family of algebras $\As$ and some of their ``correlations bimodules''. Non-commutative space-time should now be constructed topologically via the ``C*-enveloping'' of the base category of these ``bundles''  and we guess that its spectral non-commutative geometry can be recovered from the additional spectral data of the modular spectral geometries living on the total space of such ``bundles''. 

The investigation of ``(higher) categorical modular theory'' is now a priority of this program.

\end{document}